# Flight of Dynamic Targeting on the CogniSAT-6 Spacecraft

[1]Chien, S. A., [1]Zilberstein, I., [1]Candela, A., [2]Rijlaarsdam, D., [2]Hendrix, T., [2]Dunne, A., [3]Aragon, O., and [3]Miquel, J. P.

[1]steve.a.chien@jpl.nasa.gov, Jet Propulsion Laboratory, California Institute of Technology, USA

[2]Ubotica Technologies, Ireland

[3]Open Cosmos, UK

**Abstract**

Dynamic targeting (DT) is a spacecraft autonomy concept in which sensor data is acquired and rapidly analyzed and used to drive subsequent observation. We describe the low Earth orbit application of this approach in which lookahead imagery is analyzed to detect clouds, thermal anomalies, or land use cases to drive higher quality near nadir imaging. Use cases for such a capability include: cloud avoidance, storm hunting, search for planetary boundary layer events, plume study, and beyond. The DT concept requires a lookahead sensor or agility to use a primary sensor in such a mode, edge computing to analyze images rapidly onboard, and a primary followup sensor. Additionally, an inter-satellite or low latency communications link can be leveraged for cross platform tasking. We describe implementation in progress to fly DT in early 2025 on the CogniSAT-6 (Ubotica/Open Cosmos) spacecraft that launched in March 2024 on the SpaceX Transporter-10 launch.

**Introduction**

Rapid developments in New Space are enabling new paradigms for Earth Observation. Decreased launch and spacecraft costs are driving an explosion in the number of Earth Observation assets. Advances in hardware, increased risk tolerance (in part due to decreased launch and hardware costs), and nascent communications infrastructure enable new strategies for Earth Observation that leverage "edge intelligence" enabled by immediate processing of raw imagery into knowledge that can be immediately acted upon by the observing asset or rapidly transmitted to other entities for action. Intelligent networked space assets are envisioned by NASA's Earth Science Technology Office (ESTO) New Observing Strategies (NOS) [1] as well as the European Space Agency's Phi-Lab [2].

In this paper, we describe an effort to demonstrate in-space the most rapid application of edge intelligence, Dynamic Targeting [3] (DT) to the Low Earth Orbit (LEO) case. DT LEO looks ahead in the orbital track. At an orbital altitude of 500 km and a ground speed of approximately 7.5 km/s a 45 degree lookahead results in a lookahead of approximately 74 seconds before nadir [4]. Figure 1 below shows the sequence of activities in the timeline for the DT concept.

We are working to implement DT onboard the CogniSAT-6 (CS-6) [5] spacecraft. CS-6 launched in March 2024 on Transporter-10 from Vandenberg Space Force Base in California, USA. CS-6 DT has only a single imager and therefore uses the CS-6 instrument (utilizing red, green, blue, and very near infrared spectral bands) as both the lookahead and primary sensor. Onboard algorithms for analysis will use the Intel Myriad X edge processor onboard for both deep learned convolutional neural networks and for implementations of spectral analysis including: spectral angle mapping, match filters, and spectral unmixing. Several lookahead analysis techniques are planned [6,7] including: cloud detection/avoidance, thermal anomaly detection (for detection of volcanic activity or wildfires), and spectral signature detection. The CS-6 orbit allows for a 60-90 seconds lookahead lead time using a 40-50 degree lookahead. The current effort plans in-space demonstration of DT in early 2025 and we report on the status of implementation for that scheduled flight time.

DT has numerous applications. The most prevalent use case is for cloud avoidance as many instruments and science are hampered by clouds. DT for cloud avoidance is already being used in operations of the TANSO-FTS-2 sensor on the GOSAT-2 spacecraft [8] quite effectively by JAXA. In this application a lookahead visible range sensor is used to detect cloud obscuration and the primary TANSO-FTS-2 Fourier Transform Spectrometer is targeted to avoid detected clouds resulting in more usable science data. Additional studies have analyzed the productivity gains from cloud avoidance for atmospheric retrievals such as $CO_2$ and $CH_4$ [9] as well as more generic studies [3].

Mission concepts to use DT to hunt for rare science events of interest also show promise. SMICES would use DT to hunt for deep convective ice storms with a profiling radar [10]. Other concepts would use DT to target specialized radar and lidar to capture rare, transient Planetary Boundary Layer (PBL) phenomena [11].

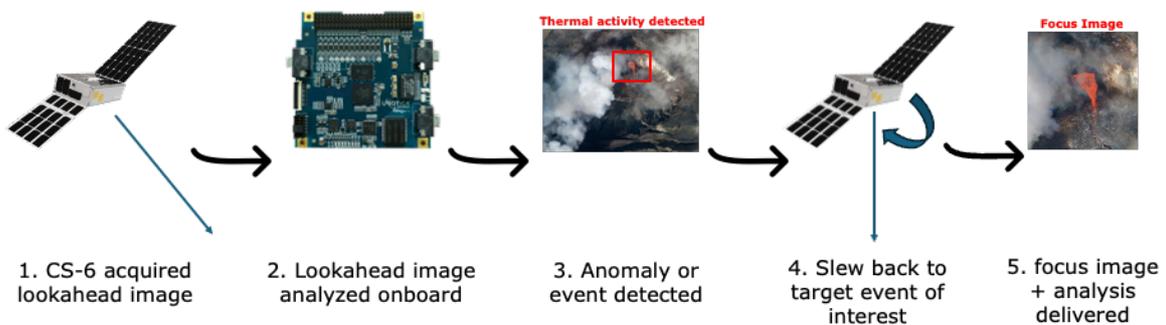

Figure 1: Overview of a dynamic targeting workflow for volcano monitoring.
*Image sources: Ubotica Technologies (steps 1, 2, & 4), Planet Labs (step 3 and 5).*

**Execution Timeline**

The challenge in implementing DT is to meet the execution timeline dictated by LEO orbital velocity[1]. As indicated, at a 500km orbital altitude, a 40, 45, and 50 degree lookahead produce timelines from lookahead imaging to nadir view of 60 to 90 seconds. Within this timeframe all of the actions indicated below 1 must be performed.

---

[1] Although alternate concepts of DT that leverage GEO and MEO data in near real-time are being explored as well.

1. acquire lookahead image; transfer image from instrument to processing unit
2. pre-process image; analyze image
3. translate analysis to targeting and instrument settings; issue reconfiguration and repointing commands
4. execute along track and across track slew
5. acquire near nadir image; analyze nadir image; downlink analysis

Many options are available in the lookahead imaging. Because CS-6 does not have a dedicated lookahead imager the primary imager does not have a large field of view. One strategy is to emulate a larger field of view coarser spatial resolution imager by smearing the pixels by executing slews across track or by reducing readout times to in effect produce larger pixels along track.

Transferring the image out of the instrument to the processing unit can take considerable time as CS-6 was not designed for the rapid instrument readout required for DT. In normal satellite operations instrument readout times of minutes are not an issue however for the DT concept readout times in the seconds are more suited for the DT timeline. In order to reduce readout time we are investigating decimation of the data (e.g. sampling only a subset of the pixels, in effect simulating a lower spatial resolution instrument), reading out only a subset of the spectral bands (or using the panchromatic band).

Also to improve the accuracy of the image analysis, we perform simple dynamic range stretching of the image (as a substitute for more sophisticated radiometric correction or conversion to top of the atmosphere radiance or reflectance).

Lookahead image analysis in principle could use almost any image analysis technique [6,7]. For practical purposes our primary three lookahead algorithms are: cloud avoidance, cloud seeking (as a SMICES or PBL surrogate), and thermal anomaly detection (as a surrogate for volcano or wildfire thermal search).

As one of the most time consuming operations in the DT timeline is the slew from lookahead to near nadir mode (this is a slew of 40-50 degrees along track), to optimize the timeline this slew can be initiated prior to completion of the lookahead image analysis. In an ideal situation, a lookahead sensor makes this slew unnecessary.

For nadir image analysis, we enable the full complement of onboard techniques we are implementing for the CS-6 sensor [6,7]. Also for a subset of the DT sequences we will also downlink the nadir analyses using the intersatellite link to highlight the rapid product delivery enabled by this technology.

**Current Status**

Flight of DT on CS-6 is currently in implementation. The onboard imagery analysis is fairly far along [6,7] and is expected to fly in Fall of 2024 as early as October 2024. Slewing tests to assess feasibility were performed on an identical spacecraft indicating that CS-6 has sufficient agility for the DT scenario using the sensor as both lookahead and nadir sensor. Timing benchmarks on meeting the hard DT requirements have been performed and indicate that there is still significant engineering work to go to meet the DT timeline. Due to the technical issues in work and the high competition for CS-6 operations time we are currently working towards an early 2025 flight demonstration of DT on CS-6.

## Conclusion

Advances in spacecraft edge computing offer tremendous potential for onboard data analysis to improve spacecraft operations. Dynamic Targeting represents an extreme edge case where a spacecraft acquires data, analyzes said data onboard, and uses that analysis to improve further remote sensing.

We describe efforts to fly DT on the CogniSAT-6 spacecraft in Low Earth Orbit. The concept has been analyzed and is in implementation for flight in early 2025. Flight demonstration of DT will pave the way for future highly autonomous Earth and deep space missions, as well as for future multi-agent intelligent, distributed observation constellations.

## Acknowledgements

Portions of this work were performed by the Jet Propulsion Laboratory, California Institute of Technology, under a contract with the National Aeronautics and Space Administration (80NM0018D0004). This work was supported by the NASA Earth Science and Technology Office (ESTO). Government sponsorship acknowledged.